\documentclass[preprint,11pt]{article}
\usepackage{amssymb}
\usepackage{amsmath}
\usepackage{graphicx}
\usepackage{epstopdf}
\usepackage[space]{grffile}
\usepackage {float}

\usepackage{authblk}

\title{A Bifurcation Monte Carlo Scheme for Rare Event Simulation}

\author[1]{Hongliang Liu \thanks{hl854@nyu.edu}}
\author[2]{Jonathan Goodman \thanks{goodman@cims.nyu.edu}}
\affil[1]{Physics Department, New York University, NY, USA}
\affil[2]{Courant Institute of Mathematical Sciences, New York University, NY, USA}

\begin{document}

\maketitle

\begin{abstract}
The bifurcation method is a way to do rare event sampling -- to estimate the probability
of events that are too rare to be found by direct simulation.
We describe the bifurcation method and use it to estimate the transition rate of a double 
well potential problem.
We show that the associated constrained path sampling problem can be addressed by a
combination of Crooks-Chandler sampling and parallel tempering and marginalization.
\end{abstract}


\section{Introduction}

There are many situations where one is interested in an event that happens rarely
on the natural time scale of the system \cite{Chemical Reaction Example 1},
\cite{Chemical Reaction Example 2}, \cite{Protein Folding Example}.
Common examples involve thermal noise assisted transitions from one potential well to another
\cite{Magnetic Switching Example}. 
Direct simulation may be impractical as a way to estimate such transition rates.

There are several approaches to do rare event simulation.
Some involve analytic or semi-analytic solution of the associated instanton or 
large deviations variational problem, e.g., \cite{Magnetic Switching Example}.
These methods apply in cases in which a typical
rare event trajectory is close to the solution of the variational problem.
(Note: ``typical'' rare event paths are very different from typical paths.)
Other methods rely less on theory, and may be preferable in situations where the
theory is unavailable or the variational problem is intractable.
Some of these, including the bifurcation method presented here, rely on the
fact that it is practical to sample the space of typical rare events by Markov chain Monte Carlo (MCMC) even
in cases where it is hard to estimate their probability.
The bifurcation method presented here is similar to the method of \cite{SungjaeJun}, which
is applied to problems in queueing theory, and to the method of \cite{{JimBeck}}, which is
applied to problems of structural reliability.  
It is similar to the method of nested sampling \cite{nested sampling}, which, in turn, is motivated by
thermodynamic integration.

In the general situation there is a random object, $\mathbf{X}$, with a probability density $f(\mathbf{X})$.
We have a function $\phi(\mathbf{X})$.
The problem is to estimate 
\begin{equation}
       p_{\alpha} = \mbox{Pr}\!\left( \phi(\mathbf{X}) > \alpha \right) \; .
\label{eq:pa} \end{equation}
We are particularly interested in these probabilities when they are very small.
In the example presented here, $\mathbf{X}$ represents a path corresponding to some stochastic
dynamics, and $\phi(\mathbf{X})$ is a scalar function of the path.
The density of the scalar diagnostic variable will be $\rho(\alpha)$, so that
\[
          \rho(\alpha)\,d\alpha 
        = \mbox{Pr}\!\left( \alpha \leq \phi(\mathbf{X}) \leq \alpha + d\alpha \right) \; ,
\]
and
\[
      p_{\alpha} = \int_{\alpha}^{\infty} \rho(\alpha^{\prime}) \,d\alpha^{\prime} \; .
\]

We give a general description of the bifurcation method here with more details for special problems in Sec. \ref{sec: Bifur for double well}.
At $k=0$, we create $N$ samples $\mathbf{X}^{(0,n)} \sim f$, for $n = 1, \ldots, N$,
and evaluate the corresponding diagnostic variables, 
\[
      \phi^{(0,n)} = \phi(\mathbf{X}^{(0,n)}) \; .
\]
These are used to estimate $\alpha_0$ as the median of the empirical distribution
\[
       \widehat{\alpha}_0 = \mbox{Median} \left\{ \phi^{(0,n)} \right\} \; .
\]
This $\widehat{\alpha}_0$ approximately bifurcates the histogram $\rho(\alpha)$ into
the upper and lower $50\%= 2^{-1}$ quantiles.
In our examples, it is possible to generate independent paths $\mathbf{X}^{(0,n)}$.
We call this process as ``\textit{pre-sampling}".

For any $\alpha$, we define the constrained probability density as
\begin{equation}
      f_{\alpha}(\mathbf{X}) = \left\{ \begin{array}{ll}
                  \frac{1}{p_{\alpha}} f(\mathbf{X})   & \mbox{ if } \phi(\mathbf{X}) > \alpha\\ [5pt]
                                      \;\;0   & \mbox{ otherwise.}
                                           \end{array} \right.
\label{eq:fa}  \end{equation}
The constrained histogram density $\rho_{\alpha}$ is similar:
\begin{equation}
      \rho_{\alpha}(\alpha^{\prime}) = \left\{ \begin{array}{ll}
            \frac{1}{p_{\alpha}} \rho(\alpha^{\prime})   & \mbox{ if } \alpha^{\prime} > \alpha\\ [5pt]
                                      \;\;0   & \mbox{ otherwise.}
                                           \end{array} \right.
\label{eq:rhoa}  \end{equation}
We assume that it is possible to use MCMC to sample $f_{\alpha}$ starting 
from a suitably generic $\mathbf{X}$ with $\phi(\mathbf{X}) > \alpha$.
In the application presented below, we may sample $f_{\alpha}$ using the Crooks-Chandler 
\cite{STPS} version of transition path sampling \cite{TPS}, see below.

The bifurcation algorithm is based on the fact that $\alpha_k$ is the median of
the $\rho_{\alpha_{k-1}}$ histogram.
At the start of step $k$, for $k \geq 1$, we have an estimate $\widehat{\alpha}_{k-1}$.
We also have a typical sample $\mathbf{X}^{(k,1)} \sim f_{\widehat{\alpha}_{k-1}}$.
We run MCMC on the distribution $f_{\widehat{\alpha}_{k-1}}$ to generate enough samples
to get a reliable estimate of the median of $\rho_{\widehat{\alpha}_{k-1}}$.
This median (the median of the samples as an estimate of the true median) is $\widehat{\alpha}_k$.
We call this process as ``\textit{bifurcation loop k}".
We also choose at random one of the samples $\mathbf{X}^{(k,n)}$ that has $\phi(\mathbf{X}^{(k,n)})>\widehat{\alpha}_k$.
This will be $\mathbf{X}^{(k+1,1)}$. Bifurcation loops end when we find the first $k$ with $\widehat{\alpha}_{k} > \alpha$, and let us denote it as $k'$. Then we can estimate $p_\alpha$ by
\begin{equation} \label{eq:estimate prob}
	\widehat{p}_\alpha = 2^{-k'} \times  \int_{\alpha}^{\infty} \rho_{\widehat{\alpha}_{k'-1}}(\alpha^{\prime}) \,d\alpha^{\prime} \; .
\end{equation}

The bifurcation method can find rare events more effectively than direct simulation.
Suppose $N= 10^7$ samples are taken at each level.
At the end of $k=60$ levels, one is getting samples of events with 
probability $p_{\alpha_{60}} = 2^{-60} \approx 10^{-18}$, having taken only $(k+1)N = 6.1\times 10^8$ samples in all.
This is possible for the following reason: given one typical sample from $f_{\alpha}$, MCMC
allows one to get many more, even when $\alpha$ is very small.

We demonstrate in this paper that the bifurcation method can be used to estimate transition
rates.
We study a system with state $X(t)$ that spends most of its time in stable region $A$ or $B$, 
with rare transitions between. We first estimate the transition probability $p_{AB}(T)$ for a given time $T$.
The random object, $\mathbf{X}$, is the path of $X(t)$ during [0, T].
The transition probability, $p_{AB}(T)$ is estimated by starting the system in region $A$ and estimating
the (small) probability that $X_T$ is in region $B$, i.e., we define the transition probability as,
\begin{equation}\label{eq: def p}
	p_{AB}(T) \equiv \langle h_B(X_T) \rangle \; .
\end{equation}
Here, $ \langle \cdots \rangle $ denotes a statistical average, and $h_B(X_T)$ is the characteristic function of region B, i.e. $h_B(X_T)=1,$ if $X_T$ $\in$ B, else, $h_B(X_T)=0$.
If $T$ is taken to be a time longer than the transient time $\tau_{\mbox{\scriptsize \em ini}}$, but much shorter than the
typical time between transitions, $\tau_{\mbox{\scriptsize \em rxn}} \approx k_{AB}^{-1}$,
the transition rate is related to the transition probability by
\begin{equation}
        Tk_{AB} \approx p_{AB}(T) \; .
\label{eq:tp} \end{equation}
For a practical computation, when $T$ is not very long, it is more accurate to estimate $k_{AB}$ as the plateau value of
\begin{equation}
	k_{AB}(t) \equiv \frac{dp_{AB}(t)}{dt} \; .
\end{equation}
It can be shown by computational examples that $k_{AB}(t)$ changes during the transient time $\tau_{ini}$. This $\tau_{\mbox{\scriptsize \em ini}}$ is related to the time needed for the system to reach its local equilibrium in region A. After that, $k_{AB}(t)$ reaches a plateau, and will be a constant for a long time $\tau_{\mbox{\scriptsize \em rxn}}$. This plateau value is the transition rate $k_{AB}$ in the empirical sense.

This approach does not require a good understanding of the system's dynamics. It only requires an order parameter $\phi(\mathbf{X})$. It is an advantage of the bifurcation method since it does not require the ''true" (or perfect) reaction coordinate (such as the committor) \cite{TPS} which is usually quite difficult or impossible to determine in advance for a complex system.

As explained above, we need an MCMC sampler for the constrained distributions
$f_{\alpha}$ given by (\ref{eq:fa}). There is a simple way to create one. 
Choose a Metropolis Hastings sampler for the unconstrained distribution $f$ that has a 
parameter, $r$, that represents a step size.
Small $r$ should correspond to proposals being close to the current sample.
Suppose the current sample, $\mathbf{X}^{(n)}$, has the distribution $f_{\alpha}$, and that
the proposal is $\mathbf{X}^{(n)} \to \mathbf{Y}$.
First perform the usual Metropolis Hastings rejection test on $\mathbf{Y}$.
If $\mathbf{Y}$ is accepted in this step, then accept if $\phi(\mathbf{Y}) > \alpha$ and set $\mathbf{X}^{(n+1)}=\mathbf{Y}$.
Otherwise, if $\phi(\mathbf{Y})<\alpha$, reject and set $\mathbf{X}^{(n+1)}=\mathbf{X}^{(n)}$.
It is easy to check that if the original method satisfies detailed balance for $f$, then
the modified method satisfies detailed balance for $f_{\alpha}$.
If $r$ is small enough and if $\phi(\mathbf{X}^{(n)}) > \alpha$, then there is a reasonable probability
that $\phi(\mathbf{Y}) > \alpha$ also.

Traditional MCMC samplers, such as the Crooks-Chandler method we use, have difficulties 
in some bifurcation loops due to the fact that paths have fluctuations on many time scales.
We propose a multi-scale version of the Crooks-Chandler method that is based on 
parallel tempering  \cite{Parallel Tempering 1} \cite{Parallel Tempering 2} and 
parallel marginalization \cite{Parallel Marginalization}.
This improves the MCMC efficiency by up to a factor of 38.

In the following sections, we begin with a brief description of the double well problem; then, we describe details of our bifurcation method for this problem, discuss the error estimation of it and methods to improve the efficiency, and describe a method to estimate the transition rate; at last, we show computational results and make a discussion.

\section{The Double Well Problem}

The transition rate in a double well potential with small noise is a well studied test problem,
which has been used to explore the effectiveness of some rare event
sampling methods \cite{Rare Events 1} \cite{Reare Events 2}.

Generally, the Langevin dynamics for a noisy system may be written as
\begin{equation} \label{eq:sde}
	\frac{dX(t)}{dt} = b(X(t))+\epsilon \dot{W}(t) \; ,  \qquad X(0) = x_0 \; , 
\end{equation}
or in the It\^{o} form 
\begin{equation} \label{eq:Ito}
	dX(t) = b(X(t)) dt + \epsilon dW(t) \; ,   \qquad X(0) = x_0 \; .   
\end{equation}
Here, $\dot{W}$ is standard white noise, and $W(t)$ is the corresponding Brownian motion. 
In a conservative force field, the drift velocity is derived from a potential, 
$b(X) = -\gamma^{-1}\triangledown U(X) $, where $\gamma$ is the friction coefficient. The parameter 
$\epsilon = \sqrt{ \frac{2k_B T}{\gamma} }$, where $k_B$ is Boltzmann's 
constant and $T$ is the temperature, determines the relative strength of thermal noise and 
deterministic dynamics.
When $U(X)$ has two wells, there are transitions from a locally stable region A to another one B. 
For small $\epsilon$, the $A \rightarrow B$ transition time is too long to be estimated by 
direct simulation. 

\begin{figure} [h]
	\begin{center}
	\includegraphics[width=80mm]{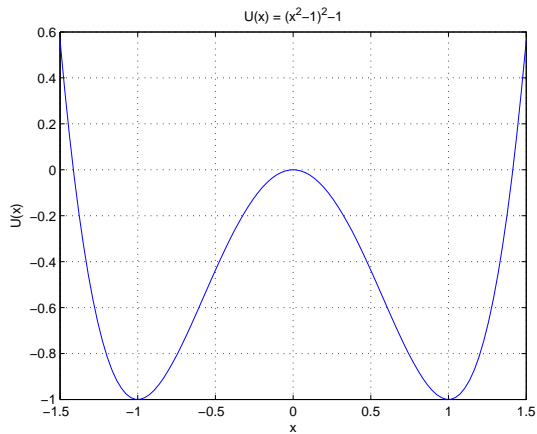}
	\caption{Double well potential}
	\label{fig:double well potential}
	\end{center}
\end{figure}

Much theoretical and computational work has gone into this rare-event problem \cite{50 years after Kramers}  \cite{Kramers theory} \cite{GL coupled fields 1} \cite{GL coupled fields 2}. 
The theoretical estimate of the transition rate is the Kramers (or Van't Hoff-Arrhenius) law, 
\begin{equation} \label{eq:Arrhenius law}
	k_{AB} = \nu \exp(-\beta E_a) \; ,
\end{equation}
where $ k_{AB} $ is the transition rate, $\beta = \frac{1}{k_B T}$, $E_a$ is the activation energy, and $\nu$ is a prefactor. 
The activation energy is
\begin{equation} \label{eq:Activation Energy}
	E_a = E_t-E_A \; ,
\end{equation}
where $E_A$ is the energy of region A, and $E_t$ is the lowest maximum energy on a path from A to B.
We take the specific potential $U(x) = (x^2-1)^2-1$, which is illustrated in 
Figure \ref{fig:double well potential}.
The A and B wells are centered about $x=-1$ and $x=1$ respectively.
Their energy is $E_A = E_B = -1$.
The transition energy is $E_t = 0$.
Therefore, the activation energy is $E_a = 1$.

\begin{figure} [h] 
  \begin{center}
  \includegraphics[width=100mm]{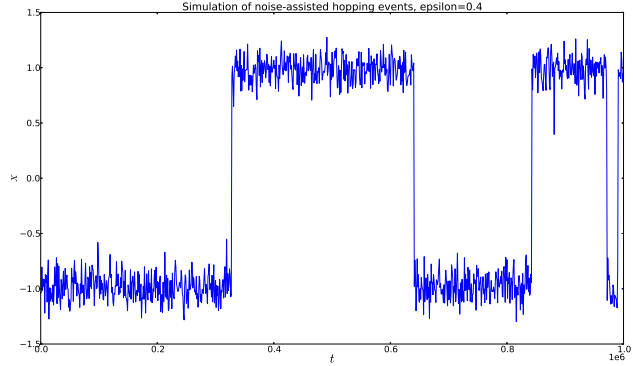}
  \caption {A path by the direct Monte Carlo simulation, with time step $\Delta t = \frac{10}{2^{10}}$. }
  \label{fig:DMC}
  \end{center}
\end{figure}

We simulate the stochastic dynamics (\ref{eq:sde}) using the Euler-Maruyama method 
\cite{Numerical SDE book}: 
\begin{equation} \label{eqn:Euler M. method}
	\begin{split}
		X_{ t_{i+1} } = X_{t_i} + b(X_{t_i})\Delta t + \epsilon \sqrt{\Delta t}Z_i \; ,\\
		t_i = i \, \Delta t \; , \qquad Z_i \sim \mathcal{N}(0,1), \mbox{ (indep)} \; .
	\end{split}	
\end{equation}
Here, $\mathcal{N}(0,1)$ is the normal distribution with mean zero and variance one.
Figure \ref{fig:DMC} gives one sample path, calculated for $\epsilon = 0.4$,  $\gamma = 1$, and $\Delta t = 10 \times 2^{-10}$, which starts from X(0) = -1.
Even with this relatively large noise, transitions are rare and isolated events.
A comparable simulation with $\epsilon = 0.2$ would show no transitions.

We wish to estimate the transition rate $k_{AB}$ when this rate is very small, 
using the bifurcation method.

\section{Bifurcation Method for the Double Well Problem} \label{sec: Bifur for double well}

We describe details of the general bifurcation method specific to the transition path problem.
The random object is a discrete path, which we denote it as $X_{[0,T]}$,
\[
	X_{[0,T]} = \{X_0, \, X_{\Delta t}, \,  X_{2\Delta t}, \, \cdots ,X_{T} \}  \; .
\]
The probability density $f(X_{[0,T]})$ is determined by (\ref{eqn:Euler M. method}),
\begin{equation}
	f(X_{[0,T]})= f(X_0) \prod_{i=0}^{T/\Delta t-1}p(X_{i\Delta t} \to X_{(i+1)\Delta t} ) \; ,
\label{eq:tp1}  \end{equation}
where $ f(X_0) = \delta (X_0-x_0)$ and
\begin{equation}
	p(X_{i\Delta t} \to X_{(i+1)\Delta t} )= \frac{1}{\sqrt{2\pi \epsilon ^2 \Delta t}}\exp[-\frac{(X_{(i+1)\Delta t } - X_{i \Delta t } -b(X_{i\Delta t }) \Delta t)^2}{2\epsilon ^2 \Delta t} ] \; .
\label{eq:tp2} \end{equation}
With this density function, we can find the statistical average of any quantity, for example,
\begin{equation}
	\langle h_B(X_T) \rangle = \int f(X_{[0,T]}) h_B(X_T) \mathcal{D}X_{[0,T]} \; .
\end{equation}
The notation 
\[
	\int \mathcal{D}X_{[0,T]} \equiv \int \prod_{i=0}^{T/\Delta t} dX_{i\Delta t}
\]
is borrowed from path integral. 

We do pre-sampling by (\ref{eqn:Euler M. method}), while in bifurcation loops,
we use the Crooks-Chandler method to sample the constrained distribution.
This method relies on the dynamics (\ref{eqn:Euler M. method}), which expresses
the path $X_{[0,T]}$ in terms of the driving noise 
\[
      Z_{[0,T]} = \{ Z_0, Z_1, \ldots \} \; .
\]
If noise $Z^{(n)}_{[0,T]}$ corresponds to $X^{(n)}_{[0,T]}$, the Crooks-Chandler method
generates an independent noise path $\widetilde{Z}_{[0,T]}$ and uses this to create a
modified noise path as
\[
         Z^{\prime}_{[0,T]} = \sqrt{1-r^2}\, Z^{(n)}_{[0,T]} + r \widetilde{Z}_{[0,T]} \; .
\]
This formula gives $Z^{\prime}_{[0,T]}$ the correct independent ${\cal N}(0,1)$ distribution.
The method then calculates $Y_{[0,T]}$, as the corresponding trajectory of 
(\ref{eqn:Euler M. method}) with noise $Z^{\prime}_{[0,T]}$.
If $Y$ is accepted, then $Z^{(n+1)}_{[0,T]}=Z^{\prime}_{[0,T]}$.
Otherwise, $Z^{(n+1)}_{[0,T]}=Z^{(n)}_{[0,T]}$.

\subsection{Efficiency and Error}
The efficiency of a Monte Carlo computation is the relation between the accuracy of 
the result to the amount of work.
Most MCMC computations have little bias, so the error is essentially determined by the 
variance of the final estimate.
The bifurcation method is a sequence of median estimates based on MCMC samples of constrained
distributions.
Therefore, the error of a bifurcation method estimate is determined by the errors in the 
median estimates.
For that reason, we review the theory of error bars for median estimates.

In a typical bifurcation step we have $N$ MCMC samples $\phi^{(1)}, \ldots, \phi^{(N)}$ that
are samples of a histogram density $\rho$.
Let $\alpha$ be the true median and $\widehat{\alpha}$ the sample median.
We wish to characterize the distribution of $\widehat{\alpha}-\alpha$.
This is done in two steps.
First we characterize the distribution of $M$, the number of samples below the true median.
If $M = \frac{1}{2}N$, then $\widehat{\alpha} = \alpha$.
The deviation $M - \frac{1}{2}N$ drives the deviation of $\widehat{\alpha}$ from $\alpha$.
The second step is to characterize this relation, which is
\[
        M - \frac{1}{2}N \approx N \cdot \left( \alpha -\widehat{\alpha} \right) \rho(\alpha) \; .
\]
There are $M$ samples at the true median and $\frac{1}{2}N$ samples at $\widehat{\alpha}$.
The right side of this relation is the number of samples between $\widehat{\alpha}$ and $\alpha$,
which is the density of samples ($\rho(\alpha))$ multiplied by the number in all ($N$) and the 
width ($\widehat{\alpha}-\alpha$). Here we ignore higher orders.
From this we derive the relation
\[
        \mbox{var}\!\left(\widehat{\alpha}\right) 
      \approx \frac{\displaystyle \mbox{var}(M)}{N^2\rho(\alpha)^2 }\; .
\]
We characterize $\mbox{var}(M)$ as described in \cite{Sokal}.
The quantity $M$ is given by
\[
      M = \sum_{n=1}^N H(\alpha - \phi^{(n)}) \; .
\]
Here, $H$ is the Heaviside function $H(\phi)=1$ if $\phi > 0$ and $H(\phi)=0$ if $\phi < 0$.
Since $\alpha$ is the median of $\phi$, the variance of each term is $\frac{1}{4}$.
The Kubo-Green variance formula is then
\[
        \mbox{var}\!\left(\widehat{\alpha}\right) 
      \approx \frac{1}{4 N_{\mbox{\scriptsize \em eff}} \, \rho(\alpha)^2 }\; .
\]
Here, $N_{\mbox{\scriptsize \em eff}} = \frac{1}{\tau} N$ is the effective sample size.
The auto-correlation time $\tau$ is the auto-correlation time of the random sequence
$H(\alpha - \phi^{(n)})$.
The central limit theorem shows that $\widehat{\alpha}$ is approximately Gaussian.
It is convenient to write this in the form
\begin{equation}
\widehat{\alpha} \approx \alpha + \frac{Z}{2 \,\rho(\alpha)\sqrt{N_{\mbox{\scriptsize \em eff}}}} \; ,
\label{eq:va}  \end{equation}
where $Z \sim {\cal N}(0,1)$.

We use this to trace the accumulation of error through the repeated bifurcation process.
We let $g(\alpha)$ be the median of the constrained distribution $\rho_{\alpha}$ of (\ref{eq:rhoa}).
The true median of the $\rho_{\widehat{\alpha}_{k-1}}$ distribution is $g_k = g(\widehat{\alpha}_{k-1})$.
The one step error formula (\ref{eq:va}) may be written
\[
          \widehat{\alpha}_k \approx g(\widehat{\alpha}_{k-1}) + 
                    \frac{\sqrt{\tau_k}}{2 \,\rho_{\widehat{\alpha}_{k-1}}(g_k)\sqrt{N_k}} \,Z_k \; .
\]
The error in $\alpha_k$ is $e_k =  \widehat{\alpha}_k- \alpha_k$, so 
$\widehat{\alpha}_k = \alpha_k + e_k$.
If $e_k$ is small, then  
\[
	g(\alpha_{k-1}+e_{k-1}) \approx g(\alpha_{k-1})+ g'(\alpha_{k-1})e_{k-1}. \\ 
\]
This gives us the error propagation relation
\[
	e_{k} \approx  g^{\prime}(\alpha_{k-1})e_{k-1} + 
	    \frac{\sqrt{\tau_k}}{2 \,\rho_{{\alpha}_{k-1}}(\alpha_k)\sqrt{N_k}} \,Z_k\; .
\]
Since the $Z_k$ are independent, this leads to the variance propagation relation
for $\sigma_k^2 = \mbox{var}(\widehat{\alpha}_k)$:
\begin{equation}
	\sigma^2_k \approx  ( g'(\alpha_{k-1}) )^2 \times \sigma_{k-1}^2 
	       + \frac{\tau_k}{ 4 \,(\rho_{\alpha_{k-1} }(\alpha_{k}))^2\,N_k } \; .
\end{equation}

Finally, we can discuss the error of the estimated probability. Suppose we want to find $p_\alpha$, and by our bifurcation algorithm we find the first $k$ with $\widehat{\alpha}_{k} > \alpha$ and denote it as $k'$.  We can write 
\[
	p_\alpha = p_{\widehat{\alpha}_{k'-1}} \times p^*_{\alpha} \; ,
\]
where
\[
	p_\alpha^{*} \equiv \int_{\alpha}^{\infty} \rho_{\widehat{\alpha}_{k'-1}}(\alpha^{\prime}) d\alpha^{\prime} \; .
\]
It can be easily shown, 
\begin{equation}
	\widehat{p}_{\widehat{\alpha}_{k'-1}} \approx 2^{-k'} + \rho(\alpha_{k'-1}) \times e_{k'-1} \; ,
\end{equation}
and
\begin{equation}
	\widehat{p}_\alpha^{*} \approx p_\alpha^{*} + \frac{Z \times \sqrt{\tau_{k'}} \sqrt{p_\alpha^{*} (1-p_\alpha^{*} ) }}{ \sqrt{N_{k'}} } \; ,
\end{equation}
where $Z \sim {\cal N}(0,1)$.
We estimate $p_\alpha$ as $ \widehat{p}_\alpha = \widehat{p}_{\widehat{\alpha}_{k'-1}} \times \widehat{p}_\alpha^{*} $. The total variance for $ \widehat{p}_\alpha $ is
\begin{equation}  \label{eqn:total error of prob}
	\mbox{var}( \widehat{p}_\alpha ) = \mbox{var}( \widehat{p}_{\widehat{\alpha}_{k'-1}} ) \times (p_\alpha^{*})^2 + \mbox{var}(	\widehat{p}_\alpha^{*}) \times ( p_{\widehat{\alpha}_{k'-1}} )^2 + \mbox{var}( \widehat{p}_{\widehat{\alpha}_{k'-1}} ) \times \mbox{var}(	\widehat{p}_\alpha^{*}) \; .
\end{equation}

Clearly $ \tau_{k} $ plays an important role in the errors. 
The first thing to increase the efficiency is to reduce $ \tau_{k} $. 
Parallel tempering, sometimes called exchange Monte Carlo \cite{Parallel Tempering 2}, 
is often used to reduce auto-correlation times. 
Parallel tempering performs MCMC on a system that consists of $L$ replicas of the original one.
This multiplies the dimension by a factor of $L$.
Traditionally, replica $l$ corresponds to a temperature $T_l$, with $l = 1,2,\cdots, L$. 
Suppose the original system is given with a temperature $T_1$, which is so low that has a 
long auto-correlation time with a traditional Metropolis-Hastings algorithm. 
The basic idea of parallel tempering is to use the fact that states can move easier in high temperatures, so we will consider L systems with temperature $T_1<T_2<\cdots<T_L$.
More generally, we can expand to L replicas with a perturbed parameter $\gamma$.
In the present application replica $l$ is a sample with potential 
\[
        U_l(x) = \frac{1}{a_l} \,U(x) \; .
\]
Larger $a_l$ corresponds to lower energy barriers.
In the present application we have chosen the $a_l$ by trial and error.

For the $L$-replica ensemble, we propose $m$ individual Crooks-Chandler moves for each
replica, and following it we propose $n$ exchange moves $\mathbf{X}_l \leftrightarrow \mathbf{X}_{l+1}$
for $l = 1, \cdots, L-1$, so we get $m+n$ samples for the original system in such moves.
According to our test, to generate N samples,  parallel tempering takes L times computational time,
when compared to the pure Crooks-Chandler method.
Therefore we must compare the $\tau$ of the simple method to $L\tau$ of parallel tempering.

\subsection{Combining Parallel Tempering with Parallel Marginalization}

The parallel marginalization method of Weare \cite{Parallel Marginalization} is similar to 
parallel tempering.
But instead of adjusting a physical parameter such as temperature, parallel marginalization 
adjusts the time step $\Delta t$.
Replica 1 is the original system.
The higher replicas have larger $\Delta t$, which we take to be $\Delta t_l = 2^{l-1}\Delta t_1$.
If $d_l$ is the dimension of the space replica $l$ lives in, then $d_l = \frac{1}{2} d_{l-1}$.
Larger time step has two consequences.
One is that an MCMC step is cheaper.
The other is that there is less dynamic range in the fluctuations.
As before, we can use the basic Crooks Chandler on any individual replica.
It allows larger $r_l$ in the basis Crooks Chandler MCMC step on the higher replicas. 
We combine parallel tempering with parallel marginalization together. So replica $l$ is a sample with potential $U_l(x)$ and time step $\Delta t_l$.

The replica exchange step is more complicated than in parallel tempering because 
$\mathbf{X}_l$ has twice as many values as $\mathbf{X}_{l+1}$.
To do an $\mathbf{X}_l \leftrightarrow \mathbf{X}_{l+1}$, we must invent new intermediate time values for $\mathbf{X}_{l+1}$
and delete intermediate time values for $\mathbf{X}_l$.
Some formalism helps describe this process.
For a given replica $l$, we write its state $\mathbf{X}_l$ as two parts $(\widehat{\mathbf{X}}_l,\widetilde{\mathbf{X}}_l)$.
The $\widehat{\mathbf{X}}_l$ part is the large time step part that can be turned into a level $l+1$ replica.
It consists of the path sampled at even numbered time steps
\[
      \widehat{\mathbf{X}}_l = ( X_{l,0}, X_{l,2\Delta t_l},\, X_{l,4\Delta t _l}, \,\cdots ) \; .
\]
This has the same number of time steps as as a whole $\mathbf{X}_{l+1}$.
The times coincide: step $2j$ of $\widetilde{\mathbf{X}}_l$ is at time $2j\Delta t_l$, which 
is the same as step $j$ of $\mathbf{X}_{l+1}$, which is $j\Delta t_{l+1}$.
It represents the slow fluctuations of the path $\mathbf{X}_l$, fluctuations on a time 
scale $2\Delta t_l$ or slower.
The $\widetilde{\mathbf{X}}_l$ part is the odd numbered time steps
\[
      \widetilde{\mathbf{X}}_l = ( X_{l,\Delta t_l},\, X_{l,3\Delta t _l}, \cdots ) \; .
\]
We think of this as representing the fine scale fluctuations in $\mathbf{X}_l$.

The replica exchange proposal has the form 
\[
       (\widehat{\mathbf{X}}_l, \widetilde{\mathbf{X}}_l,\mathbf{X}_{l+1}) \;\to\; (\mathbf{X}_{l+1}, \widetilde{\mathbf{Y}}_l, \widehat{\mathbf{X}}_l) \; .
\]
The proposed new $\widehat{\mathbf{X}}_l$ is $\mathbf{X}_{l+1}$, the proposed new $\mathbf{X}_{l+1}$ is $\widehat{\mathbf{X}}_l$, and the proposed new $\widetilde{\mathbf{X}}_l$ is $\widetilde{\mathbf{Y}}_l$,
which is a sort of stochastic interpolation 
from the values of $\mathbf{X}_{l+1}$ to the odd numbered time steps with time step $\Delta t_l$. 
In the present application, we generate a Gaussian  random path $\{\zeta_j\}$ 
with independent components $\zeta_j \sim \mathcal{N} (0,\frac{\epsilon^2 \Delta t_l}{2})$
and $j = 0, 1, \cdots, T/\Delta t_{l+1} -1$. For each $j$, let
\[
	\widetilde{Y}_{l,j}  = \zeta_j + 0.5( X_{l+1, j\Delta t_{l+1}} + X_{l+1, (j+1) \Delta t_{l+1}} ) \; ,
\]
i.e. the proposal density of $\widetilde{\mathbf{Y}}_l$ is
\[
	q_l(\widetilde{\mathbf{Y}}_l|\mathbf{X}_{l+1}) \propto \exp( \sum_{j} - \frac{( \widetilde{Y}_{l,j}- 0.5( X_{l+1, j\Delta t_{l+1}} + X_{l+1, (j+1) \Delta t_{l+1}} ) )^2}{ \epsilon ^2 \Delta t_l } ) \; .
\]

To be more precisely in mathematics, in the exchange move between replicas $l$ and $l+1$,
denote the current state as $x$, and the proposed new state as $y$, i.e.
\[
    x \equiv (\mathbf{X}_1, \cdots, \widehat{\mathbf{X}}_l, \widetilde{\mathbf{X}}_l,\mathbf{X}_{l+1} \cdots) \; , \qquad
	y \equiv (\mathbf{Y}_1, \cdots, \widehat{\mathbf{Y}}_l, \widetilde{\mathbf{Y}}_l,\mathbf{Y}_{l+1}, \cdots ) \; .
\]
Detailed balance requires that
\[
	f(x)Q(x\to y)A(x,y) = f(y)Q(y \to x)A(y,x) \; .
\]
Here
\[
	f(x) = f_1(\mathbf{X}_1) \cdots f_l(\widehat{\mathbf{X}}_l, \widetilde{\mathbf{X}}_l)f_{l+1}(\mathbf{X}_{l+1})\cdots f_L(\mathbf{X}_L)\; ,
\]
with $f_l$ have the Euler Mayurama distributions (\ref{eq:tp1}) and (\ref{eq:tp2}).
(Just be careful that different $l$ has different potential and time step.) $f(y)$ is similar.
The proposal density from $x \to y$ is
\[
	Q(x\to y) = q_l( \widetilde{\mathbf{Y}}_l|\mathbf{X}_{l+1} ) \delta(\widehat{\mathbf{Y}}_l-\mathbf{X}_{l+1} ) \delta(\mathbf{Y}_{l+1}-\widehat{\mathbf{X}}_l) \prod_{l'\neq l \, \text{and} \, l'\neq l+1}\delta(\mathbf{Y}_{l'}-\mathbf{X}_{l'}) \; ,
\]
and $A(x,y)$ is the probability that we accept the proposal $x\to y$.
We can get
\begin{equation}\label{eqn:GPPM acceptance}
	A(x,y) = \mbox{min} ( 1, \frac{ q_l(\widetilde{\mathbf{X}}_l|\widehat{\mathbf{X}}_l) f_l(\mathbf{X}_{l+1},\widetilde{\mathbf{Y}}_l) f_{l+1}(\widehat{\mathbf{X}}_l) }{ q_l(\widetilde{\mathbf{Y}}_l|\mathbf{X}_{l+1}) f_l( \widehat{\mathbf{X}}_l, \widetilde{\mathbf{X}}_l) f_{l+1}(\mathbf{X}_{l+1}) } ) \; .
\end{equation}

Similar to parallel tempering, we propose $m$ individual Crooks-Chandler moves for each
replica $l$ followed by $n$ replica exchange moves. Since the size of the $(l+1)_{th}$ replica is 
only half of  the $l_{th}$ replica, it takes less work than parallel tempering.
According to our test, to generate N samples,  parallel tempering and marginalization takes 2 times computational time,
when compared to the basic Crooks-Chandler method.
Therefore we should compare the $\tau$ of the simple method to $2\tau$ of parallel tempering and marginalization.

\subsection{Calculating the Transition Rate} \label{sec: cal transition rate}

By our bifurcation method, we can find the transition probability  $p_{AB}(t)$; we can repeat the process for different $t$; at last, we can estimate $k_{AB}(t)$ by $k_{AB}(t) \approx \frac{ p_{AB}(t+\Delta t) - p_{AB}(t)}{\Delta t}$. However, this takes too much time. 

There is a more convenient and quicker way to estimate $k_{AB}(t)$, which is described in \cite{TPS}. We explain its brief idea here. Since we initialize X(0) in region A, the description can be simplified. We write $p_{AB}(t)$ as 
\[
	p_{AB}(t)= p_{AB}(t')\times R(t,t') \; ,
\]
where $ R(t,t') \equiv  \frac{p_{AB}(t)}{p_{AB}(t')}$. 
According to our definition of the transition probability (\ref{eq: def p}),
\begin{equation}
	 R(t,t') = \frac{ \langle h_B(X_t) \rangle }{ \langle h_B(X_{t'} ) \rangle } \; .
\end{equation}
Consider a trajectory $X[0, t'']$ with $t''\ge t, t'$.
Define the path function $H_B(X_{[0,t'']})$ which is unity if at least one state along the trajectory $X_{[0,t'']}$ is within B and vanishes otherwise, i.e.
\[
	H_B(X_{[0,t'']}) \equiv \max_{0 \leq s \leq t''} h_B(X_s) \; .
\]
Define
\[
	\langle h_B(X_t) \rangle _{t''}^* \equiv \frac{ \int \mathcal{D}X_{[0,t'']} f(X_{[0,t'']})h_B(X_t)H_B(X_{[0,t'']}) }{ \int \mathcal{D}X_{[0,t'']} f(X_{[0,t'']})H_B(X_{[0,t'']}) } \; .
\]
For $0 \leq t \leq t''$, $ h_B(X_t)H_B(X_{[0,t'']}) = h_B(X_t) $. $ \forall \, t, t'$,  if $ 0 \leq t, t' \leq t''$, it can be easily shown that
\begin{equation}
	R(t,t') = \frac{ \langle h_B(X_t) \rangle _{t''}^* }{ \langle h_B(X_{t'}) \rangle _{t''}^* } \; .
\end{equation}
$ \forall \, t \in [0, t'']$,  $\langle h_B(X_t) \rangle _{t''}^*$ can be determined in a  single transition path sampling run,
in which the transition path ensemble is a set including all trajectories which start from A and visit B in $[0, t'']$ and the acceptance probability in the Crooks-Chandler method is $H_B(X_{[0,t'']})$.
So we can easily get $R(t,t')$.

Now we can describe procedures to quickly estimate $k_{AB}(t)$. First, for a fixed time $t'$, we use our bifurcation method to estimate $p_{AB}(t'$).  Next, for a chosen $t''$ ($t'' \ge t'$), starting from a successful trajectory, $ \forall \, t $, $0 \leq t \leq t''$, $R(t, t')$ is determined  from a single transition path sampling run. 
Combining these two steps, by the following equations,
\begin{eqnarray}
	p_{AB}(t) &=& p_{AB}(t') \times \frac{\langle h_B(X_t) \rangle_{t''}^* }{\langle h_B(X_{t'}) \rangle_{t''}^* } \;,\\
	k_{AB}(t) &\equiv& \frac{dp_{AB}(t)}{dt} = \frac{d \langle h_B(X_t) \rangle_{t''}^*  }{dt} \times \frac{p_{AB}(t')}{\langle h_B(X_{t'}) \rangle_{t''}^* } \; ,
\end{eqnarray} 
we can easily estimate $k_{AB}(t)$.

\section{Computational Examples}
Taking the double well potential problem as our computational example, we want to know the transition rate for X changing from well A (near -1) to well B (near 1).

\subsection{Improvement of Efficiency by Parallel Tempering and Marginalization }\label{sec: example of 0.4}
First, we show the advantage of combining parallel tempering and marginalization to the Crooks-Chandler method. We use the {\em acor} software of \cite{Acor} to estimate the auto-correlation time.

We set our simulation as following: $\epsilon = 0.4$, $\gamma = 1$, $ X(0) = -1 $, $T = 10$, $\Delta t$ of the original system is set to be $ \frac{10}{2^{10}}$, and in bifurcation loops, $ r = 0.4 $. We simply choose $\phi(X_{[0,T]}) = X(T)$, and characterize $X(10)>0.5$ as one successful transition.

\begin{figure} 
  \begin{center}
  \includegraphics[width=100mm]{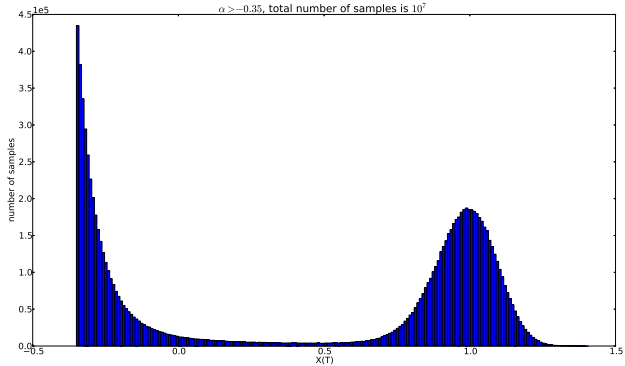}
  \caption {Constrained histogram of X(10) for the double well potential problem, when $\epsilon = 0.4$ . }
  \label{fig:hist}
  \end{center}
\end{figure}

When applying the basic Crooks-Chandler method in our bifurcation, we find that in some bifurcation loops the auto-correlation time is pretty long. Take sampling conditional on $ X(10)>-0.35$ as an example. We make a histogram of $X(10)$, Fig. \ref{fig:hist}. It shows that in our interested sample space, $X(10)$ has a multi-modal distribution with peak regions separated far away. In such case it is quite often that samples generated by the Crooks-Chandler method have a pretty long auto-correlation time. With $N=10^8$ samples, we estimate $ \tau \approx 50,000 $.

When parallel tempering is applied, with the ratio of Crooks-Chandler moves and exchange moves set to be 2:1,  the best result we get is using $L = 5$ and $\{ a_l \} = \{1, 1.2, 1.5, 2.0, 3.0 \}$. With $N=10^7$ samples, we estimate $\tau \approx 550$ when sampling conditional on $ X(10)>-0.35$. Compared with the pure Crooks-Chandler method, we make it more efficient by a factor of $\frac{50,000}{5\times 550} \approx 18$.

When the parallel tempering and marginalization method is combined to the Crooks-Chandler algorithm, the efficiency can be improved more. We take Crooks-Chandler moves and exchange moves with ratio 2:1. With $N=10^7$ samples, by setting $L = 6$, $\{ a_l \} = \{1, 1.4, 1.8, 2.2, 2.6, 3.0 \}$, we get $\tau \approx 650$ when sampling conditional on $ X(10)>-0.35$. Compared with the pure Crooks-Chandler method, we make it more efficient by a factor of $\frac{50,000}{2\times 650} \approx 38$.

\subsection{Estimation of a Small Transition Rate}
We use our methods to estimate the transition rate when $\epsilon = 0.2$, $\gamma = 1$. We set $ X(0) = -1 $, $\Delta t$ of the original system is set to be $ \frac{5}{2^{9}}$, and in bifurcation loops, $ r = 0.2 $. We combine the Crooks-Chandler algorithm with parallel tempering and marginalization, using $L = 6$, $\{ a_l \} = \{1, 1.4, 1.8, 2.2, 2.6, 3.0 \}$. The ratio of Crooks-Chandler moves and exchange moves is set to be 2:1.

\subsubsection{Estimate the Transition Probability for a Given Time}
\begin{figure} 
  \begin{center}
  \includegraphics[width=140mm]{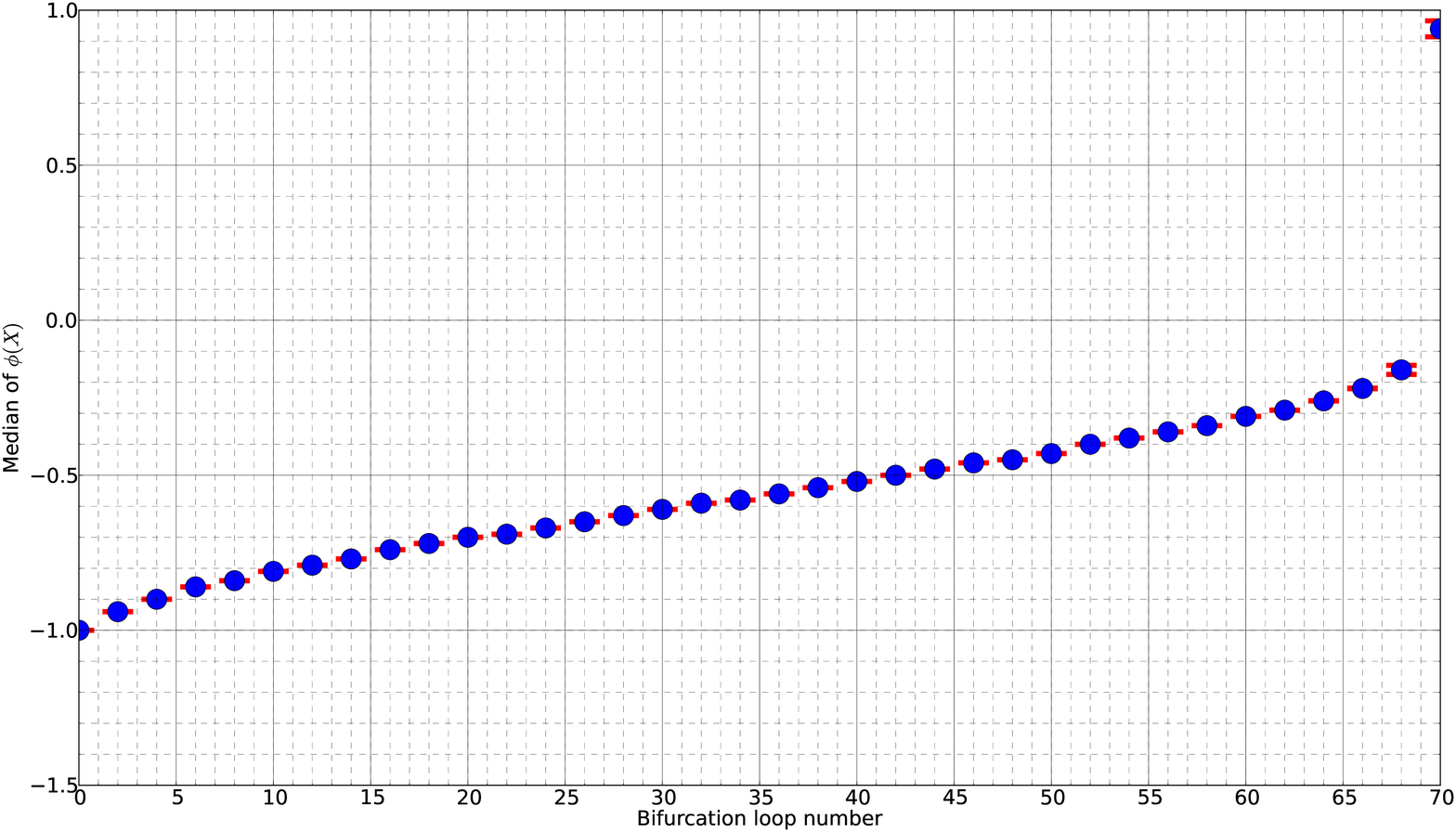}
  \caption {Bifurcation sampling result for the double well potential problem, when $\epsilon = 0.2$ . }
  \label{fig:B_error}
  \end{center}
\end{figure}

For a fixed time $t' = 5$, we use our bifurcation method to estimate the transition probability. We choose $\phi(X_{[0,t']}) = X(t')$ and characterize $X(5)>0.5$ as one successful transition.  Setting the number of samples in each bifurcation loop to be $N_k = 10^7$, we get Fig. \ref{fig:B_error}. In it, the x-axis stands for the bifurcation loop number k, to make the error bar clear we only plot with k = even numbers, and k = 0 means the pre-sampling; y-axis stands for the median of $X(t')$ in the bifurcation loop $k$. We estimate $p_{AB} (5)$ as $\widehat{p}_{AB} (5) = (5.8 \pm 1.8 )\times 10^{-22}$.

\subsubsection{Estimate the Transition Rate} 
We use the quick method described in Sec. \ref{sec: cal transition rate}. To decide $R(t,t')$, we choose $t'' = 10$,  the sample space is the collection of paths which have reached region B in the period [0, 10], and $N = 4\times 10^8$.

\begin{figure} 
  \begin{center}
  \includegraphics[width=100mm]{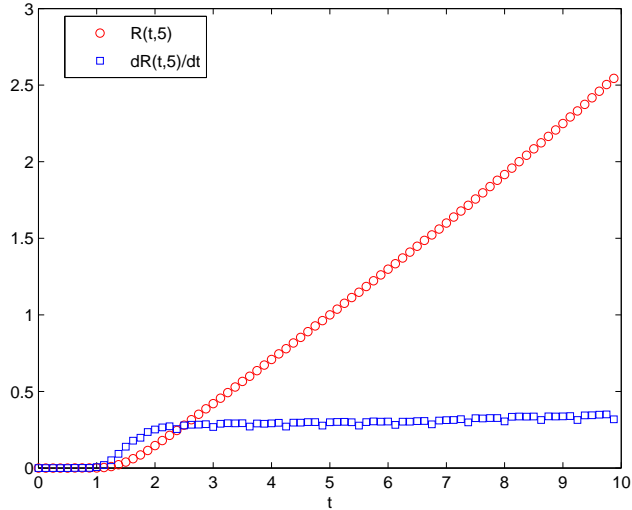}
  \caption { the ratio R(t, t') and its derivative with respect to t, when $\epsilon = 0.2$ }
  \label{fig:TPS}
  \end{center}
\end{figure}

Our result is shown in Fig. \ref{fig:TPS}.
In this figure, x-axis stands for time, and y-axis stands for $ R(t,5) $ in red line or $ \frac{d R(t,5) }{dt} $ in blue line. So we estimate $ \widehat{k}_{AB} \approx 0.3 \times \widehat{p}_{AB}(5) \approx (1.7 \pm 0.5)\times 10^{-22}$. In Fig. \ref{fig:TPS}, we can see that after a short transient time $\tau_{\mbox{\scriptsize \em ini}}$, which is about 2, $k_{AB}(t)$ reaches a plateau.

\section{Conclusion and Discussion}

In this paper, we propose a bifurcation method to estimate the transition probability, and apply it to the double well potential problem. In each bifurcation loop, the Crooks-Chandler algorithm is used to do MCMC sampling. We find that during our bifurcation process, some bifurcation loops have very long auto-correlation times when only using the Crooks-Chandler method. When $\epsilon =0.4$ and $\gamma = 1$, it happens when we do sampling conditional on $X(10)>\alpha$, where $ \alpha \in [-0.5, 0.5]$. Sparked by the parallel tempering method and the parallel marginalization method, we propose the ``parallel tempering and  marginalization" method. With $\epsilon =0.4$ and $\gamma = 1$, we take sampling conditional on $X(10)>-0.35$ as a typical example. Our method reduces the time needed to get an almost independent sample from 50,000 units to 1,300 units, while parallel tempering only reduces it to 2750 units. Pointed out by the renormalization theory, judicious elimination of variables by renormalization can reduce long range spatial correlations \cite{Renormalization}. So in some other problems, it can reduce the computing time even more. This is another possible advantage to combine the marginalization method together with parallel tempering.

With $\epsilon = 0.2$, $\gamma = 1$, we test our whole algorithm. By about $7.5\times 10^8$ samples, we get the transition probability for $T=5$ is $\widehat{p} (X(5)>0.5) = (5.8 \pm 1.8) \times 10^{-22}$. Following that, combining Crooks-Chandler and parallel tempering and  marginalization, by one transition path sampling run in the collection of paths which have reached region B in the period [0, 10], with $N = 4\times 10^8$ samples, we get $\widehat{k}_{AB} \approx (1.7 \pm 0.5 )\times 10^{-22}$. Analytically, by the Kramers' theory, the transition rate is
\begin{equation}
	k_{AB}=\frac{\omega_0\omega_1}{2 \pi \gamma}\exp(-\frac{2\Delta U}{\gamma \epsilon^2}) \; ,
\end{equation}
where $ \omega_0 = \sqrt{ \frac{d^2U}{dx^2}|_{x = -1} } $, $ \omega_1 = \sqrt{ -\frac{d^2U}{dx^2}|_{x = 0} } $, and $ \Delta U = U(0) - U(-1),$
so $k_{AB, \mbox{\scriptsize \em Kramer}} =  \frac{2\sqrt{2}}{\pi}e^{-\frac{2}{0.04}} = 1.7 \times 10^{-22}$. For such a small transition rate, we get a pretty good estimation of the transition rate within affordable computational time by our bifurcation method.

At last, we would like to mention that the bifurcation method can be generalized. In the bifurcation loop $k$, we can set $\alpha_k$ to satisfy the following equation,
\begin{equation}
	\mbox{Pr}_{\alpha_{k-1}}(\phi(X_{[0,T]})\geq \alpha_k) \equiv \int_{\phi(X_{[0,T]})\geq \alpha_k }f_{\alpha_{k-1}}(X_{[0,T]})\mathcal{D}X_{[0,T]} = c \; ,
\end{equation}
where $0< c <1$. If c is 0.5, it is our bifurcation method proposed in this paper.

\end{document}